# Singularity Blockchain Key Management

*Sumit Vohra*
Singularity, Singapore

*Abstract—* web3 wallets are key to managing user identity on blockchain. The main purpose of a web3 wallet application is to manage the private key for the user and provide an interface to interact with the blockchain. The key management scheme ( KMS )  used by the wallet to store and recover the private key can be either custodial, where the keys are permissioned and in custody of the wallet provider or non-custodial where the keys are in custody of the user. The existing non-custodial key management schemes tend to offset the burden of storing and recovering the key entirely on the user by asking them to remember seed-phrases. This creates onboarding hassles for the user and introduces the risk that the user may lose their assets if they forget or lose their seed-phrase/private key.  In this paper, we propose a novel method of backing up user keys using a non-custodial key management technique that allows users to save and recover a backup of their private key using any independent sign-in method such as google-oAuth or other 3P oAuth.

I.                      INTRODUCTION

The advent of the decentralized internet has facilitated a new generation of exciting benefits to people who buy, possess, or otherwise engage with digital products. Public blockchains represent the foundation of the decentralized web, empowering users from anywhere in the world to manage and maintain ownership of these digitally native assets. Users can enjoy the digital and IRL benefits and use cases associated with ownership of such products or they can transfer ownership with just a few clicks.

Web3 wallets, designed to interact with decentralized applications and blockchain networks, present unique challenges that make their usage complex for the average user. This introduction explores the reasons why using Web3 wallets can be difficult and the barriers that hinder widespread adoption. The transition from Web2 to Web3 introduces unfamiliar concepts, including cryptographic key management, gas fees, and decentralized identity.

These complexities often overwhelm users accustomed to centralized platforms and traditional financial systems. The technical nature of web3 wallets, which require users to handle private keys securely and understand blockchain operations, presents a significant learning curve that hampers their adoption.

Furthermore, the fragmented nature of web3 wallets exacerbates the problem. With numerous wallet providers, each with its own user interface, features, and compatibility with different blockchains, users face confusion and fragmentation when selecting and switching between wallets. The lack of standardisation in user experiences and wallet functionalities adds to the complexity and hinders seamless integration into daily digital activities.

    Another challenge is the management of multiple blockchain networks and assets. Unlike traditional wallets that primarily deal with a single currency, web3 wallets need to support various cryptocurrencies and tokens across different blockchains. This multi-chain environment introduces interoperability challenges, requiring users to navigate different networks, interfaces, and protocols to access and manage their assets effectively.

Moreover, security concerns pose a significant hurdle. While web3 wallets offer greater control and ownership of assets, they also place the burden of security on the users themselves. The risk of losing private keys, falling victim to phishing attacks, or interacting with malicious smart contracts increases the potential for financial loss. Users must adopt stringent security practices, such as hardware wallets or cold storage, to mitigate these risks effectively.

A familiar, web2 style sign-in based key backup solution can greatly contribute to user adoption of decentralized applications and web3 technologies by simplifying the onboarding process and enhancing user convenience. Here's how:

I. **Streamlined Onboarding:** Traditional web3 wallets require users to generate and securely manage cryptographic keys, which can be a technical and intimidating process for non-technical users. However, if an integrated and simplified option of backing up keys is made available, users can simply opt for it as part of the onboarding flow, and don't need to indulge in technically complex methods to manage their keys. Users can also leverage their existing accounts authentication methods, eliminating the need to create and manage new credentials. This streamlined onboarding process reduces friction and lowers the barrier to entry, making it easier for users to get started with web3 applications.

II. **Enhanced Convenience:** By integrating familiar sign-in functionality, users can access their web3 wallet with a single click, eliminating the need to remember and enter complex passwords or cryptographic keys. Leveraging their existing social media accounts for authentication provides a seamless and familiar experience, increasing user convenience and encouraging broader adoption.

III. **Trust and Familiarity:** Social media platforms have already established a level of trust and familiarity with users. By integrating popular Single Sign-On capabilities to auto backup and restore user keys, web3 wallets can leverage this trust and familiar environment, easing user concerns related to security and privacy. Users are more likely to adopt web3 applications when they can associate them with the same authentication methods they use daily, reducing skepticism and promoting confidence in the technology.

II.                   THE WALLET DILEMMA

Any web3 wallet needs to have security and convenience as two properties to be a perfect solution for an average user. It should have high security for a user to trust it with their valued digital assets. It must also have the intuitiveness for an average consumer to easily understand and onboard onto the wallet. This creates what we call the wallet dilemma, where security and convenience have an inverse relation. Though a solution like storing your wallet's key in a cold wallet might be good from security perspective, it's also hard for the user to manage this external overhead with an a big risk of storing and losing this storage device. Also using such device every single time to sign a transaction is a big convenience hurdle. On the other hand, solutions like custodial wallets though easy to use, create large security loopholes where potentially untrusted 3rd parties have constant access to user keys. A perfect balance of two is hard to strike as both are equally important for onboarding next billion users to web3.

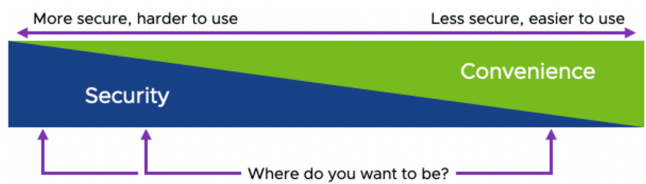

1. **Security**: Security is of utmost importance for web3 wallets. Web3 wallets are designed to interact with decentralized applications (dApps) and manage users' digital assets, such as cryptocurrencies and non-fungible tokens (NFTs), in a decentralized manner. As a result, they are highly targeted by malicious actors due to the potential for financial gain. Here are a few reasons why security is crucial for web3 wallets:

    i. **Asset Protection**: Web3 wallets manage the private keys or seed phrases that grant access to users' funds. If these credentials are compromised, unauthorised individuals can gain control over the assets, leading to financial losses. Strong security measures help safeguard these sensitive details.

    ii. **Decentralization:** Web3 wallets aim to eliminate the need for intermediaries and put users in control of their assets. However, this also means that users bear full responsibility for their wallet security. The absence of central authorities makes it even more critical to have robust security practices.

    iii. **Irreversibility**: Blockchain transactions are typically irreversible, meaning that once a transfer is made, it cannot be undone. If a malicious actor gains access to a user's web3 wallet and initiates unauthorised transactions, it may not be possible to recover the lost funds. Emphasising security helps prevent such situations.

    iv. **Phishing and Malware Attacks**: Phishing attacks and malware targeting web3 wallet users are prevalent. Hackers use deceptive techniques to trick users into revealing their private keys or installing malicious software that compromises their wallets.

2. **Convenience:** Convenience plays a significant role in the overall user experience and adoption of decentralized applications. Here are a few reasons why convenience matters:
    i. **Ease of Use**: The convenience of using a web3 wallet can significantly impact its adoption. If the process of setting up and using a web3 wallet is overly complex or time-consuming, it may discourage users from engaging with decentralized applications. By offering a seamless and user-friendly experience, web3 wallets can attract more users to the ecosystem.

    ii. **Cross-platform Access**: Convenience also extends to the ability to access web3 wallets across different platforms and devices. Users should be able to seamlessly switch between desktop and mobile versions of their wallets, ensuring access to their digital assets regardless of the device they are using. This flexibility enables users to manage their assets on the go and enhances convenience.

    iii. **Recovery**: Users can easily misplace, forget or lose their private keys and seed phrases. An easy, integrated solution to securely recover backed-up user keys would greatly benefit users in making their interactions with web3 effortless.

A close to ideal solution would comprise of just enough friction to create a secure enough environment for a user to trust their web3 wallet for securing it's transactions and assets and at the same time don't feel overburdened by the steps required to secure and use it.

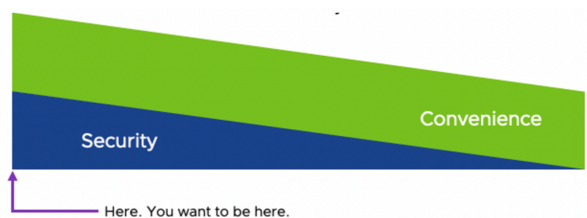

A. *Current State of Web3 Wallets*

Current web3 wallets can be classified into two main categories custodial and non-custodial wallets.

*1)* **Custodial Wallets**: Custodial wallets, also known as hosted wallets, refer to a type of cryptocurrency wallets where a third-party service provider holds and manages the private keys on behalf of the user. In this arrangement, the user's cryptocurrency holdings are stored and controlled by the custodial wallet service.

With custodial wallets, users typically create an account with the wallet service and entrust their private keys to the provider. The wallet service is responsible for generating and securing the private keys, as well as managing the associated public addresses and transactions on behalf of the user.

Custodial wallets offer a user-friendly experience as users do not need to worry about the technicalities of key management, backups, or security measures. They often provide additional features such as user interfaces, customer support, and integration with various platforms and exchanges.

However, custodial wallets come with certain trade-offs. Since the wallet service has control over the private keys, users must place all their trust in the service provider to properly secure their funds and handle transactions as instructed. If the custodial wallet service experiences a security breach or becomes inaccessible, there is a risk of potential loss or compromise of the user's funds.

Examples: Most Centralized Exchanges manage their user wallets in a custodial manner wherein the user funds are managed by the exchange using a custodial service like Bitgo [1] or Fireblocks [2]

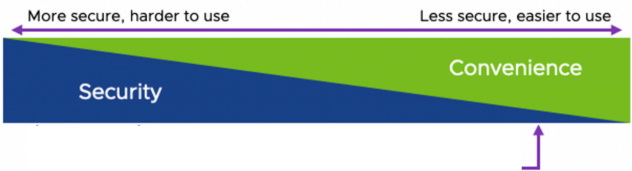

2) **Non-Custodial Wallets**: A non-custodial wallet, also known as a self-custody wallet or user-controlled wallet, is a type of cryptocurrency wallet where the user has full control and ownership of their private keys and funds.

Key management for a non-custodial wallet can be done either through a self-security model or a shared security model.

   a) **Self-security model:** In a self-security model, the user is solely responsibly for security and management of their wallet's private key. This requires users to save their seed-phrases in secured digital or physical storage. Examples of such wallets include Metamask[3], Coinbase Wallet[4], Trust Wallet[5]. Though these wallets offer superior self-custodial structure, where the wallet control is in the hands of the user, there is a big risk involved if the user forgets their seed phrase or looses their primary device. Also the interoperability between multiple platforms becomes harder, for example a game built on Unreal Engine has a hard time integrating with a user's Metamask wallet. Interoperability on a blockchain level is also hard, since a Solana seed phrase would be different from an Ethereum seed phrase. Onboarding to these wallets is also a big hassle as it involves users to create additional passwords and remember lengthy seed phrases. There are also Smart contract based wallets like Argent[6] that ease the recovery part by giving an option to create multiple custodians but at the same time increases the number of seed phrases to safeguard. and severely limits web3 operations and interoperability.

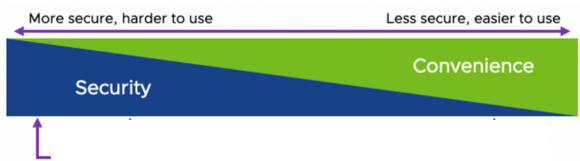

   b) **Shared Security Model:** In this model, the private keys are sharded and distributed across multiple independent parties. At the time of signing a transaction, either the private key is generated with Shamir secret sharing [6] based models like ReedSolomon Erasure Encoding [7] or MPC [8] based architectures where the signature is generated without the key generation. Examples of this include o-auth 2.0 based key management schemes where user's existing sign-in, such as Google OAuth, is used in combination with the user's device to store private key information. Web3auth [9] using torus network [10] is one such example. We believe that, if done right, this construct can provide sufficient security and at the same time can be used to create an easy onboarding experience for the end user. In the next section we discuss various Shared Security Models, their pros and cons and why we believe none of the currently popular models are optimal.

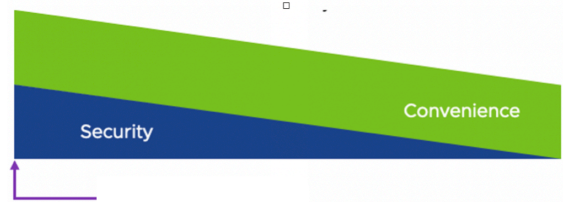

B. *Shared Security Models*

This section covers in-depth analysis of social sign-in based shared security models. This model opens up to novel key management systems where web3 wallets can leverage the security of traditional web2 OAuth2.0 structures like "sign in with Google" or "sign in with Facebook" and combine it with non-custodial schemes to achieve superior security and ease of use thereby helping user adoption.

1) **Blockchain based models (e.g. web3Auth)**: In this model, key management solutions use a private blockchain network to shard and distribute the private keys of the user. This network is a private decentralized network (~10 nodes) that uses threshold based cryptography to generate the private key once the threshold number of nodes authenticate the user using OAuth2.0. The scheme preserves the non-custodial nature of key creation where no single party controls the user key. However, this creates high-latency as every login has to go through consensus layer of the blockchain which results in high login and transaction signing latency of O(~10s). Moreover, since blockchains store data permanently, if the security of the chain is somehow compromised, the entire user key is exposed and at risk of being stolen at scale even without requiring the user's login credentials. It can also be argued that if more than 5 validators are compromised then the user's keys can be stolen by an internal attack. Web3Auth is a prominent example of a provider using this technique.

2) **HSM storage based models (e.g. Magic Link and Ginco)**: This option relies on using a relayer alongside AWS KMS and HSM storage to create access-tokens for users to access their private keys. While the HSM storage itself may be secure, the centralized relayer used to access it can be compromised to reveal the access token of the desired user. Any corruption or deletion of the HSM resource which is being actively paid by the provider can result in a total loss of user keys forever. Also AWS KMS requires the provider to store a delegated key, and only allows signing operations within their infra that makes exporting the user's seed phrase or private-key theoretically impossible, unless the seed phrase or key is also being stored separately. For these reasons, this solution can be interpreted as custodial in nature [15].

3) **MPC based models (e.g. web3Auth MPC wallet):** In this model, there is no key generated. Instead, multiple parties are enlisted that need to sign every transaction and the signatures would only be done if the user satisfies certain conditions, such as successfully signing into an authentication engine. Though superior in terms of security severely limits the interoperability of their wallet as there are no seed phrases or private keys and the user can never move their wallet to a different wallet provider. Also, if any of the signatories cease to exist or cannot be reached, the user's wallet may never be recoverable. Some providers add MFA based encryption but this results in additional interoperability cost as now private key gets tied to a particular device and loss of device or change in device results in loss of control over the user's address.

4) **Smart Contract /Account Abstraction based models (e.g. Sequence wallet):** Smart contract based wallets create a new smart contract per user, which is the actual store of tokens and is accessed through one or more "EOA wallets" (i.e. regular blockchain addresses). Wallet providers choose how the user can access the contract for transaction signing or EOA change. For example, some providers allow the user to create multiple EOA wallets that each have access to the smart contract, to create redundancy in case keys of one EOA are lost. Such structures, though non-custodial, clearly don't solve UX complexity. Most popular providers try to solve such UX issues by requiring the provider's EOA to be one of the signatories of each transaction as well as logging in to the wallet, provided the user passes some o-auth (or other) web2 authentication. In all such implementations, we have found that the provider essentially becomes a gatekeeper to the user's access to their smart wallet and since there is no way for the user to extract the smart contract's seed phrase, the user can never move their wallet to a different provider. Hence, such solutions can easily be construed as custodial in nature. From a usability perspective, smart contract wallets cost gas fee to create, have latency due to multi-signatory structure and are confusing for an average user as they must understand how the smart contract and EOA addresses interact.

To summarise, a few different solutions for key management exist, however they all make certain trade-offs that either make them custodial in nature, or negatively impact latency, interoperability, security, or all of the above.

III. SINGULARITY KEY MANAGEMENT

***Introduction:*** Singularity is a non-custodial, web2 sign-in based crypto wallet key management solution. It uses a novel, shared-security pattern to guard the backup of user's private keys.

***Motivation:*** As mentioned above, bringing 100s of millions of users to use web3 apps will require significant simplification of user onboarding while maintaining the benefits of decentralization and security provided by web3. Even though some techniques for key management exist, they have several tradeoffs like high latency, security vulnerabilities or lack of interoperability which limit their adoption. Custodial alternatives can be dangerous in the long term if a potential exploit gets executed leading to permanent tampering of user's trust on such wallets. At Singularity, we want to create a truly non-custodial key backup solution that can be operated by the user using existing web2 sign-in methods, and at the same time, is fast, secure, interoperable with other wallets and works in all types of environments that web3 apps are being built in.

***Key Concepts:*** These are a few concepts that need to be explained before we move to Singularity's architecture.

1) ***Distributed Key Generation:***

    DKG is a cryptographic process in which multiple parties contribute to the calculation of a shared public and private key set. The participation of a threshold of honest parties determines whether a key pair can be computed successfully. Distributed key generation prevents single parties from having access to a private key. The involvement of many parties requires Distributed key generation to ensure secrecy in the presence of malicious contributions to the key calculation. [11]

    Singularity uses a double sided DKG, combining the async DKG process with a runtime DKG on the client side to generate what we call a master entropy. The process is explained later on in this paper.

2) ***Torus Network:***
    This network is a private decentralised network ( ~10 nodes ) that uses threshold based cryptography to generate the private key once the threshold number of nodes authenticate using oauth2.0.[16]

3) ***OAuth2.0 and Open-Id Connect:***

    OAuth 2.0 and OpenID Connect (OIDC) are authorisation and authentication protocols respectively.

    OAuth2.0 is a authorisation framework used to grant limited access to resources without sharing user credentials. It focuses on delegated access and managing access tokens.

    OIDC builds on top of OAuth 2.0 and adds an authentication layer. It provides identity services and user authentication capabilities in addition to authorisation.OIDC enables Single Sign-On by allowing user once to authenticate with an OpenID Connect provider and then access multiple applications without re-authentication.

    OIDC specifies a standardised way for client applications to interact with an identity provider. This makes it easier for developers to integrate authentication and identity services across various platforms and services.

    Singularity and Torus network both use social IDP providers to authenticate a user. The user is uniquely

identified by their (verifier_url, verifier_id). verifier_url is generated by the IDP (e.g. Google) for each provider when they register. verifier_id is a unique identifier for the user like sub. [14]

4) *Elliptic curve encryption :*

To encrypt a message for a recipient, their public key is used. In Elliptic Curve Cryptography (ECC), a public-private key pair is generated and used for various cryptographic operations, such as encryption, decryption, digital signatures, and key exchange. Here's how the public-private key pair works in ECC:
   a. The message is transformed into a point on the elliptic curve using a suitable encoding scheme.
   b. A random value is chosen, and the sender performs scalar multiplication of the recipient's public key with this random value.
   c. The resulting point is combined with the encoded message point using an elliptic curve point addition operation.
   d. The combined point is transformed back into a cipher-text, which is sent to the recipient.
   e. The recipient can decrypt the cipher-text using their private key.
   f. They perform scalar multiplication of their private key with the received point.
   g. The resulting point is subtracted from the combined point to obtain the encoded message point.
   h. The encoded message point is decoded to retrieve the original message.

5) *Cookie and Session Management:*

Cookies are the most common form of storing user session information. This can include storing access tokens which are passed to server for authentication before returning resource info. Cookies are the most compatible in terms of browser support and platform support.
Singularity uses a novel combination of http-only cookie and local storage as a primary mode of storage for keeping the encrypted device share. This will be explained later in this paper.

6) *Shamir Secret Sharing :*

Shamir's secret sharing is a cryptographic algorithm that allows a secret to be divided into multiple shares, which are distributed among different participants. The secret can only be reconstructed when a sufficient number of shares are combined together. This technique provides a form of redundancy and can be used to protect sensitive information.
   a. The secret is represented as a polynomial of degree k-1, where k is the minimum number of shares required to reconstruct the secret.
   b. Each participant is assigned a share, which corresponds to the value of the polynomial at a specific point.
   c. To reconstruct the secret, at least k shares are needed.
   d. By using Lagrange polynomial interpolation, the original secret can be reconstructed from the available shares.

7) *Lagrange Polynomial Interpolation:*
Lagrange interpolation is a mathematical technique used to find a polynomial function that passes through a given set of data points. It allows us to estimate the values between the given data points based on the polynomial interpolation. The Lagrange interpolation formula calculates the polynomial function as a linear combination of Lagrange basis polynomials. Given a set of data points (x_0, y_0), (x_1, y_1), ..., (x_n, y_n), the Lagrange interpolation polynomial function is defined as

$$P(x) = \sum_{i=0}^{n} y_i \cdot L_i(x)$$

   i. P(x) is the interpolated polynomial function.
   ii. n is the degree of the polynomial, which is equal to the number of data points minus one.
   iii. y_i represents the y-coordinate of the i-th data point.
   iv. L_i(x) is the i-th Lagrange basis polynomial, defined as:

$$L_i(x) = \prod_{j=0, j \neq i}^{n} \frac{x - x_j}{x_i - x_j}$$

   v. In the above formula, x_i represents the x-coordinate of the i-th data point.
   vi. The Lagrange interpolation formula allows us to find the polynomial function that passes through the given data points and can be used to estimate the value of the function at any point within the range defined by the data points.

*Singularity KMS:* Singularity Key management System is core to our backup solution and works on the following principles.

1) *Distributed Key Generation:* Singularity Distributed key generation process uses double sided DKG for key generation process. This is done so that no single party has logical control over the key generation. We use double sided DKG where in the entropy to generate the private key is generated by the Torus Network running an async DKG fork on its individual nodes[12]. The entropy generated from this process is utilised in a sync-DKG process on the client side where we use server-entropy and user's device entropy to generate what we call a master entropy.

$$\text{Master Entropy} = \text{syncDKG}(\text{ServerEntropy}, \text{DeviceEntropy}, \text{asyncDKG}(\text{Torus Network}))$$

You can assume master entropy to be a combination of randomness from Torus Network, device and Singularity's server.

2) **Distributed Key Storage:** Singularity distributed key storage process makes sure that keys are sharded and held in **3 independent storages**. We use Torus Network, user's device and Singularity server as three independent storage layers to create 3 separate shards: a) Device shard on the user's device, b) Torus shard on the Torus Network and c) Singularity shard on Singularity server. Since our key generation process requires 2/3 shards, an impact on one storage has no impact on other and key generation happens successfully. Also we have hack-resistance of one storage, so if one storage gets compromised, we have sufficient time to rotate the other storage keys and replace the compromised storage.

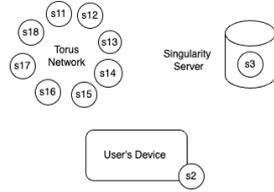

Inside Torus Network, the user's shard is further split into 9 shards to make sure no individual node owner has complete access to even one shard of our private key.

3) **Encryption at rest:** Singularity servers, user's device and Torus Network all hold the shards in encrypted format. Singularity server shards are dynamically encrypted with a salt that's on an auto-rotate policy.

4) **Recovery Mechanism:** Singularity KMS has recovery mechanism in case the user loses or changes their device. From the user's perspective they would have to just login with the same social-sign-in provider on the new device and the Singularity recovery client software will recover the user's keys from Torus Network and Singularity servers, re-generate a new device shard for the user's new device and store it in their new device.

5) **Revocability mechanism:** Singularity KMS implements a novel revocability mechanism, that enables our private key shards to be replaced with new shards in case our server shards / Torus shards or device shards are compromised.

$$E\,k\,p \leftarrow \text{Generate\_Elliptic\_Key\_Pair}(\text{rand}())$$
$$E\,1, E\,2, E\,3 \leftarrow \text{SHAMIR\_2/3\_SPLIT}(E\,k\,p)$$
$$s\,1 \rightarrow E(s\,1, E\,k\,p) \rightarrow s\,1' \rightarrow \text{torus\_storage}$$
$$s\,2 \rightarrow E(s\,2, E\,k\,p) \rightarrow s\,2' \rightarrow \text{device\_storage}$$
$$s\,3 \rightarrow E(s\,3, E\,k\,p) \rightarrow s\,3' \rightarrow \text{server\_storage}$$
$$E\,1 \rightarrow \text{torus\_storage}$$
$$E\,2 \rightarrow \text{device\_storage}$$
$$E\,3 \rightarrow \text{server\_storage}$$

Singularity KMS generates Complimentary Elliptic key pair on the user's device at the time of private key generation. This key pair is used to encrypt s1, s2 and s3 (private key shards generated by our DKG algorithm mentioned above). The key pair is then split into 3 shards using Shamir splitting and saved into three independent storages.

To decrypt the shards we use the following algorithm.

$$E\,2 \leftarrow \text{device\_storage}$$
$$E\,3 \leftarrow \text{server\_storage}$$
$$E\,k\,p \leftarrow \text{Shamir\_Lagrange\_Generate}(E\,2, E\,3)$$
$$s\,1 \leftarrow \text{DECRYPT}(s\,1', E\,k\,p)$$
$$s\,2 \leftarrow \text{DECRYPT}(s\,2', E\,k\,p)$$
$$\text{private\_key} \leftarrow \text{Shamir\_Generate}(s\,1, s\,2)$$

If any of the (s1',e1),(s2',e2), or (s3',e3) are compromised we can regenerate Ekp, decrypt old key and encrypt it with new Ekp'.

The following algorithm will be used to revoke an existing encryption key and shard and replace it with new encryption key and new shard.

$$E\,1 \leftarrow \text{torus\_storage}$$
$$E\,2 \leftarrow \text{device\_storage}$$
$$E\,3 \leftarrow \text{server\_storage}$$
$$E\,k\,p \leftarrow \text{Shamir\_Lagrange\_Generate}(E\,2, E\,3)$$
$$E\,k\,p' \leftarrow \text{Generate\_Elliptic\_Key\_Pair}(\text{rand}())$$
$$s\,1 \leftarrow \text{DECRYPT}(s\,1', E\,k\,p)$$
$$s\,2 \leftarrow \text{DECRYPT}(s\,2', E\,k\,p)$$
$$s\,3 \leftarrow \text{DECRYPT}(s\,3', E\,k\,p)$$
$$s\,1 \rightarrow E(s\,1, E\,k\,p') \rightarrow s\,1'' \rightarrow \text{torus\_storage}$$
$$s\,2 \rightarrow E(s\,2, E\,k\,p') \rightarrow s\,2'' \rightarrow \text{server\_storage}$$
$$s\,3 \rightarrow E(s\,3, E\,k\,p') \rightarrow s\,3'' \rightarrow \text{device\_storage}$$
$$E\,1' \rightarrow \text{torus\_storage}$$
$$E\,2' \rightarrow \text{device\_storage}$$
$$E\,3' \rightarrow \text{server\_storage}$$

6) **Low Latency Cost:** To reduce our latency and still maintain non-custodial nature of our key backup, we use the Singularity server shard and the user's device shard for all transaction signing, after the user logs in the first time. This separates our solution from providers that use a pure blockchain based model, which have to suffer node latencies owing to threshold querying of the blockchain network shards on each login. Transaction signing latency is reduced from O(10s) to O(100ms) through this novel approach.

$$E\,2 \leftarrow \text{device\_storage}$$
$$E\,3 \leftarrow \text{server\_storage}$$
$$E\,k\,p \leftarrow \text{Shamir\_Lagrange\_Generate}(E\,2, E\,3)$$
$$s\,1 \leftarrow \text{DECRYPT}(s\,1', E\,k\,p)$$
$$s\,2 \leftarrow \text{DECRYPT}(s\,2', E\,k\,p)$$
$$\text{private\_key} \leftarrow \text{Shamir\_Lagrange\_Generate}(s\,1, s\,2)$$

7) **Platform & Blockchain Interoperable:** Singularity KMS was designed to be interoperable in terms of both the blockchain as well as the application platform. Our device dependency is managed through iframe to store the device share in an origin separated from the hosted application. To achieve this we manage the device share through an encrypted cookie and local storage via an externally injected iframe. On the blockchain level, we use the master entropy to derive BIP standards for each individual blockchain, thereby making our solution cross-chain from day-1.

8) **Disaster management:** Singularity KMS is designed to be death proof, where in if Singularity servers our dead, users would be able to recover their seed phrases through Torus Network and their device. To achieve this we maintain a death route linked to IPFS decentralized UI client which can seamless combine the two shards in case Singularity servers are offline

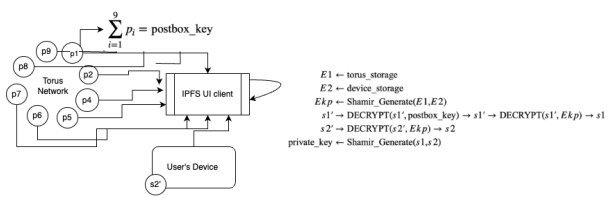

in a disaster event and users can recover their seed-phrases/private key.

9) **Wallet Interoperability:** Singularity KMS enables for seed phrase generations which is not possible in many other solutions. This creates cross wallet interoperability as the user can take their export their seed phrase and migrate to another wallet of their choice (Metamask, hardware wallet, etc).

10) **UI Independence:** Singularity's key backup solution can be used in a completely headless manner and the integrating application can own the entire user experience, select the SSO provider and manage all other details.

### *Singularity Key Generation Flow on sign-up:*

Singularity key generation process happens at runtime on the user's device during their first sign in. Once the user is verified by the SSO provider, we create a private key for the user using the following process:

$$E k p \leftarrow \text{Generate\_Elliptic\_Key\_Pair}(\text{rand}())$$
$$E1, E2, E3 \leftarrow \text{SHAMIR\_SPLIT}(Ekp)$$
$$s1 \leftarrow \text{randomEntropy1(device)}$$
$$\text{asyncDKG(Torus Nodes)} \rightarrow \text{postboxkey} \rightarrow \text{Enc}(s1, \text{postboxkey})$$
$$\rightarrow s1' \rightarrow \text{Enc}(s1', Ekp) \rightarrow s1' \rightarrow \text{torus\_storage}$$
$$\text{randomEntropy(server)} \rightarrow s2 \rightarrow \text{Enc}(s2, Ekp) \rightarrow s2' \rightarrow \text{server\_storage}$$
$$\text{randomEntropy2(device)} \rightarrow s3 \rightarrow \text{Enc}(s3, Ekp) \rightarrow s3' \rightarrow \text{device\_storage}$$
$$\text{syncDKG\_Shamir\_Lagrange\_generate}(s1, s2, s3) \rightarrow \text{private\_key}$$
$$E1 \rightarrow \text{torus\_storage}$$
$$E2 \rightarrow \text{device\_storage}$$
$$E3 \rightarrow \text{server\_storage}$$

**Torus Storage**
S1 is generated using a key called postbox key derived from async DKG[13]. In async DKG, we get already generated shards for a particular (verifier, verifier_id) from threshold nodes of torus network. These shards are combined using Lagrange polynomial interpolation on the user's device to generate postbox key. Postbox key is used to encrypt the randomEntropy1 output to generate s1'. s1' is further encrypted with Ekp to generate s1'. s1' is stored on the torus network storage layer. Members of the Torus Network are validator nodes that operate the Distributed Key Generation, Proactive Secret Sharing and Key Assignment protocol, and consist of geographically distributed and diverse businesses institutions.
In no particular order, current node operators are: ENS, Binance, Etherscan, Polygon, Ziliqa, Tendermint, Ontology, SKALE, Torus.

**Singularity Server Storage**
S2 is generated on client side using random entropy generated from singularity's server. S2 is encrypted using Ekp and stored inside a private subnet of AWS-Aurora cluster. This is further encrypted with a rotating salt stored in AWS-secret manager.

**Device Storage**
S3 is generated using random-entropy on client side. randomEntropy2 is random function composed of vanilla randomJS function, mixed with client's device parameters like IP,etc. This is encrypted using Ekp to generate s3'. On the user's device we store s3' in the http-only cookie making it inaccessible to any kind of javascript. The cookie is domain marked with s9y.gg (Singularity's domain) and an IPFS domain . This along with externalising the origin using iframe makes sure that the encrypted device shard is inaccessible for any kind of XSS and CSRF attacks. The E3 is stored in the local storage of the browser instance.

### *Singularity Key Generation Flow on sign-in:*

Singularity Key generation flow involves fetching E2,S2' and E3,S3' from device and server storage. Since we don't have to fetch from multiple nodes of Torus network in each consecutive login, the whole process of follow-up signing is ultra low latency.

$$E2 \leftarrow \text{device\_storage}$$
$$E3 \leftarrow \text{server\_storage}$$
$$Ekp \leftarrow \text{Shamir\_Generate}(E2, E3)$$
$$s1 \leftarrow \text{DECRYPT}(s1', Ekp)$$
$$s2 \leftarrow \text{DECRYPT}(s2', Ekp)$$
$$\text{private\_key} \leftarrow \text{Shamir\_Generate}(s1, s2)$$

**Ekp Fetching**
E2 is stored in the local storage of user's domain isolated iframe. E3 is fetched from the server where the SSO identity token is verified before the E3 is returned. For additional protection, all server storage is encrypted at rest with a rotating secret. Once both E2 and E3 are present on the user's device. Ekp is generated inside a special web-worker process.

**Device Shard Fetching**
Device shard is fetched using a special web-worker process running inside a domain isolated iframe. This web-worker process further guarantees of no JS manipulation while fetching the device shard. The encrypted shard goes through our domain which verifies the user SSO identity token, recovers the cookie and puts in the response and send the encrypted s1' back. This s1' is decrypted using Ekp on the user's device.

Another thing to note here is that if the user changes the device, we can use the torus shard to regenerate the private key. We further go ahead and regenerate a new device shard for this new device and follow the same flow as above for every subsequence login.

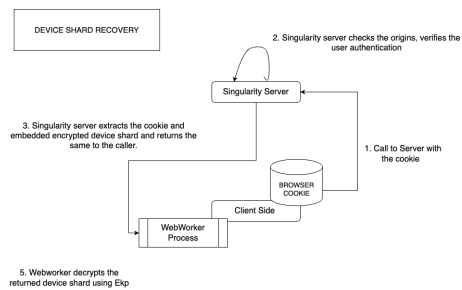

**Server Shard Fetching**
Server shard is fetched from the server, after the SSO identity token is verified.

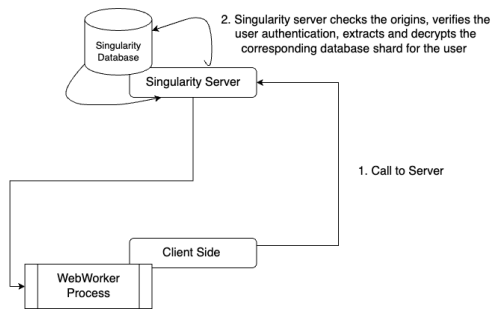

## IV. Summary

In summary, Singularity Key Management optimizes the security vs convenience trade off by providing the best in class key security for users, but at the same time, support a familiar, web2 style sign in method for users to effortlessly backup and restore their private keys. The Singularity solution solves for several of the limitations in the Shared Security key management models widely available today. By sharding the keys into 3 separate independent data stores with independent user verification and access, the Singularity solution is non-custodial. By using a state of the art cookie based device shard management approach, the Singularity solution protects users from malicious cross-scripting hacks. By using the novel Distributed Key Generation, revocability algorithm and IPFS based recover flows, the Singularity solution allows for intervention and re-recovery in case of disaster scenarios like hacks or Singularity servers becoming inaccessible. And by adopting Shamir Secret Sharing, the Singularity solution creates an interoperable key storage, that allows users the complete flexibility of moving their key to any wallet app they desire.